\documentclass[twocolumn,trackchanges]{aastex631}
\usepackage{amsmath}

\newcommand\lsun{L_{\odot}}
\newcommand\msun{M_{\odot}}

\begin{document}

\title{Considering the Single and Binary Origins of the Type IIP SN~2017eaw}
\correspondingauthor{K. Azalee Bostroem}
\email{bostroem@arizona.edu}

\author[0000-0002-4924-444X]{K.\ Azalee Bostroem}
\affiliation{Steward Observatory, University of Arizona, 933 North Cherry Avenue, Tucson, AZ 85721-0065, USA}
\altaffiliation{LSSTC Catalyst Fellow}
\author[0000-0002-7464-498X]{Emmanouil Zapartas}
\affiliation{Institute of Astrophysics, FORTH, N. Plastira 100,  Heraklion, 70013, Greece}
\affiliation{IAASARS, National Observatory of Athens, Vas. Pavlou and I. Metaxa, Penteli, 15236, Greece}
\author[0000-0001-5530-2872]{Brad Koplitz}
\affiliation{School of Earth \& Space Exploration, Arizona State University, 781 Terrace Mall, Tempe, AZ 85287, USA}
\author[0000-0002-7502-0597]{Benjamin F. Williams}
\affiliation{Department of Astronomy, University of Washington, 3910 15th Avenue NE, Seattle, WA 98195-0002, USA}
\author[0000-0002-6440-1087]{Debby Tran}
\affiliation{Department of Astronomy, University of Washington, 3910 15th Avenue NE, Seattle, WA 98195-0002, USA}
\author[0000-0001-8416-4093]{Andrew Dolphin}
\affiliation{Steward Observatory, University of Arizona, 933 North Cherry Avenue, Tucson, AZ 85721-0065, USA}
\affiliation{Raytheon Technologies, 1151 East Hermans Road, Tucson, AZ 85706, USA}

%%%%%%%%%%%%%%%%%%%%%%%%%%%%%%%%%%%%%%%%%%
\begin{abstract}
%%%%%%%%%%%%%%%%%%%%%%%%%%%%%%%%%%%%%%%%%%
Current population synthesis modeling suggests that 30-50\% of Type II supernovae originate from binary progenitors, however, the identification of a binary progenitor is challenging. 
One indicator of a binary progenitor is that the surrounding stellar population is too old to contain a massive single star.
Measurements of the progenitor mass of SN~2017eaw are starkly divided between observations made temporally close to core-collapse which show a progenitor mass of 13-15 $\msun$ (final helium core mass $M_{\rm He,core}=4.4-6.0~\msun$ - which is a more informative property than initial mass) and those from the stellar population surrounding the SN which find $M\leq10.8~\msun{}$ ($M_{\rm He,core}\le3.4~\msun$). 
In this paper, we reanalyze the surrounding stellar population with improved astrometry and photometry, finding a median age of $16.8^{+3.2}_{-1.0}$ Myr for all stars younger than 50 Myr ($M_{\rm He,core}=4.7\msun$) and $85.9^{+3.2}_{-6.5}$ Myr for %all 
stars younger than 150 Myr.
$16.8$ Myr is now consistent with the helium core mass range derived from the temporally near explosion observations for single stars. 
Applying the combined constraints to population synthesis models, we determine that the probability of the progenitor of SN~2017eaw being an initially single-star is 65\% compared to 35\% for prior binary interaction. 
$85.9$ Myr is inconsistent with any formation scenarios.  
We demonstrate that combining progenitor age constraints with helium core mass estimates from red supergiant SED modeling, 
late-time spectra, and indirectly from light curve modeling 
can help to differentiate single and binary progenitor scenarios and provide a framework for the application of this technique to future observations.

\end{abstract}


%%%%%%%%%%%%%%%%%%%%%%%%%%%%%%%%%%%%%%%%%%
\section{Introduction} \label{sec:intro}
%%%%%%%%%%%%%%%%%%%%%%%%%%%%%%%%%%%%%%%%%%
In the canonical picture of single-star evolution, massive stars (M$\gtrsim$8 $\msun{}$) are thought to end their lives as core-collapse supernovae (SNe).
These stars and their explosive deaths strongly influence galaxy evolution through ionizing flux, chemical enrichment, energy feedback, and star formation   \citep[e.g.][]{2020kobayashi,2014vogelsberger,2015wang,2018hopkins}.
In the single-star paradigm, the hydrogen envelope of the most massive stars is removed through strong stellar winds which increase with progenitor mass \citep[see][and references therein]{2014smith}.
It is therefore expected that those SNe that show hydrogen in their spectra originate from relatively lower mass massive stars ($M<25-30 ~\msun$), the majority of which become red supergiants (RSG) prior to explosion \citep[e.g.,][]{Heger+2003}.

This picture is complicated by the fact that $>$50\% of O and B stars are expected to form in binary systems \citep{2012sana,Kobulnicky+2014,Moe+2017}. 
Although the effect of binary mass stripping to hydrogen-poor SN is widely accepted \citep[e.g.,][]{Eldridge+2013}, it should also play a significant role for hydrogen-rich SNe  \citep[e.g.,][]{Podsiadlowski+1992,Podsiadlowski1992,Justham+2014}. 
In fact, \citet{Zapartas+2019} find that 30-50\% of the progenitors of hydrogen-rich SNe interacted with a companion over their lifetimes with the majority of these interactions either gaining mass through Roche-lobe overflow or through mergers. 
Although the possible effects of these binary paths, such as their role in the diversity of Type II SNe \citep[e.g.,][]{Eldridge+2018} or the explosion of stars with initial masses less than 8 $\msun$ \citep{Zapartas+2017a}, are being explored, the majority of the current literature still focuses on single-star origins for these events.

While it is clear that binary evolution affects the progenitors of hydrogen-rich SNe, it is challenging to identify individual SNe which have been affected by binary evolution.  
This is because the observable effects are often similar to those that could be explained by single-star evolution.
If binary mass accretion or merging occurred early enough in the evolution of a Type II SN progenitor, 
no circumstellar material due to that interaction is expected to linger in the vicinity during the SN. 
Similarly, the extra angular momentum of a spun up binary accretor \citep{de-Mink+2013} is lost when the progenitor expands to become a RSG.
In addition, although merging may form a stellar product with an unusual core to envelope mass ratio compared to single stars, the position in the HR diagram of the RSG that the product eventually becomes is largely unaffected \citep{2020farrell}. 
Finally, the light curves of hydrogen-rich SNe are easily explained by both single and binary progenitors \citep{2018eldridge, 2018morozova}.

Although it is hard to determine the binary origin of a progenitor star from a single observable, it is possible that the combination of observables can paint a more complete picture. Binary mass gainers and mergers can form higher core masses than the stars of the same initial mass \citep[e.g.,][]{Podsiadlowski+1992,Justham+2014} allowing stars with initial masses below $\sim$8 $\msun{}$ to explode.
As the evolution of these binary products will still be dictated by the long timescales of their initial masses, \citet{Zapartas+2021} predicted the existence of luminous hydrogen-rich RSG progenitors, with high preSN core masses, in older stellar populations than expected from single-star  evolution. They thus suggested this discrepancy among different observational methods about the inferred progenitor properties (i.e. its initial mass) could be an indirect but clear signal of the binary history of a Type II progenitor. 

SN~2017eaw exploded in NGC 6946, colloquially known as the Fireworks Galaxy due to its high number of SNe and was classified as a Type IIP SN \citep{2017Xiang}. 
At $<$ 10 Mpc, a multitude of techniques were used to measure its zero age main sequence (ZAMS) mass including pre-explosion observations with the Hubble Space Telescope (HST) and Spitzer Space Telescope of the RSG progenitor \citep{2019vandyk, 2018kilpatrick, 2019rui}, light curve modeling \citep{2020morozova, 2020goldberg, 2019szalai}, late time spectra modeling \citep{2019szalai, 2019vandyk}, and analysis of the star formation history of its stellar neighborhood \citep{2018williams, 2021koplitz}.
The comparison of these masses is striking: for all methods which evaluate mass based on observations of the star at the end of it's life, a progenitor of 13-15 $\msun{}$ is found.
However, both \citet{2018williams} and \citet{2021koplitz} failed to find evidence of young ($\lesssim$~33 Myr) stars present in the immediate vicinity, placing the progenitor mass at $\leq 10.8$$\msun{}$.
Based on the significant discrepancy of these inferred properties of the progenitor of the Type IIP SN~2017eaw, \citet{Zapartas+2021} pointed this source as the best candidate for a progenitor of a hydrogen-rich SN with history of binary mass exchange.

In this paper we re-analyze all of these measurements, with particular attention to the age (and therefore initial mass) derived from the stellar populations surrounding SN~2017eaw. 
In \autoref{sec:final_he_core}, we explain our rationale for using final He core mass rather than ZAMS mass throughout this paper.
We present an overview progenitor masses presented in the literature, derive He core masses from these ZAMS masses, and detail assumptions used throughout the paperabout distance and extinction in \autoref{sec:data}. 
In \autoref{sec:modeling} we discuss previous binary modeling and \autoref{sec:reanalysis}  presents a new analysis of the age of the stellar population surrounding the SN site.
We use binary evolution models to interpret these results in \autoref{sec:discussion}, setting tight constraints on the nature of the progenitor. 
Finally, \autoref{sec:conclusion} summarizes our results.

%------------------------
\section{Final Helium Core Mass of 2017eaw}\label{sec:final_he_core}
%------------------------
Although initial (i.e. ZAMS) mass is typically used to characterize supernova progenitors, in our analysis we focus on the final helium core mass (i.e. preSN core mass; $M_{He, core}$) rather than initial mass for a number reasons. Firstly, for SN progenitors that experience binary mass accretion or merging, the initial mass loses its meaning and its significance as a predictive evolutionary property for the final state of the progenitor. In addition, the direct progenitor imaging and the nebular spectrum observational techniques directly probe the preSN core mass instead of the (arguably more intuitive) initial mass through the preSN luminosity which is tightly correlated to the core mass, e.g. see Fig 2 of \citealt{2009smartt} or the oxygen abundance \citealt{Jerskstand+2014}. And finally, it reduces the dependency of our results on the uncertain mass-loss rates of these stars and the uncertainties in physical processes inside the star during its evolution (e.g., convective overshooting). We thus argue that the final helium core mass is a more appropriate property to characterize the Type II SN progenitors, being applicable even for binary progenitors \citep{Zapartas+2021}. 
For the remainder of this paper the final helium core mass should be considered the primary description of progenitor mass and we provide the ZAMS mass only to facilitate comparison to the literature.

%%%%%%%%%%%%%%%%%%%%%%%%%%%%%%%%%%%%%%%%%%
\section{Methods}\label{sec:data}
%%%%%%%%%%%%%%%%%%%%%%%%%%%%%%%%%%%%%%
Our study combines many different analyses that place various constraints on the possible progenitor of SN~2017eaw all using single star evolutionary models. 
Observations that originate temporally close to explosion directly measure the final state of the progenitor from which an initial mass is inferred. 
Alternatively, calculating the age of the stellar population at the SN site directly probes the age (and therefore initial mass) of the progenitor system. 
In theory, if the SN and its surrounding stellar population are modeled correctly (and the assumption of non-binarity is correct), these methods should agree.
In this section we review assumptions about metalicity, distance, and extinction as well as both the initial and final mass measurements in the literature.
Masses are summarized in \autoref{tab:finalmass}.
From each initial mass we give a final He core mass based on the models used to infer the ZAMS mass.
Finally, for all He core masses, we derive a self consistent ZAMS mass using \autoref{eq:helium_zams_singles} in this paper.

%------------------------
\begin{table*}[]
    \centering
    \begin{tabular}{|c|c|c|c|}
    \toprule
    Final Helium  & Initial Mass [$\msun$]  & Method &  Reference\\
    core mass [$\msun$]& reported  (derived from final He core) &&\\
    \hline
    4.4 & 15 (13.4) & SN light curve &  \citealt{2020morozova}\\ %sukhbold 2016
    $4.6 ^{a}$, 6.8, 7.7 & $14 ^{a}$ (13.9), 19 (19.5), 22 (21.7) & SN light curve &   \citealt{2020goldberg}\\
    $4.8$ &   $13\pm^{4}_{2}$  (14.5) & RSG SED &   \citealt{2018kilpatrick}\\
    $5.1$ &  $13-15$  (15.3) & RSG SED &   \citealt{2019vandyk}\\
    $6.0$ &  $12\pm2$ (17.6) & RSG SED &   \citealt{2019rui}\\
    4.4 &  15 (13.4) & SN late-time Spectra &  \citealt{2019szalai}\\%Heger & Woosley 2007 as found in Sukhbold 2016
    4.4 &  15 (13.4) & SN late-time Spectra & \citealt{2019vandyk}\\%Heger & Woosley 2007 as found in Sukhbold 2016
    \hline
    $\leq 3.4^{b}$ & $ \leq 10.8$ (10.9) & Population age & \citealt{2018williams, 2021koplitz}\\
    $4.7^{b}$ & $14$ (14.3) & Population age & This study\\
    \botrule
    \end{tabular}
    \caption{The final helium core mass of 2017eaw is shown in column 1, based on the observational method in column 3 and its reference at column 4. The reported initial mass at that study (and the  initial mass derived from the He core mass  based on \autoref{eq:helium_zams_singles}) are shown in column 2.  
     For comparison we show the initial mass value inferred from population age methods, discussed in \autoref{sec:popConstrain}. 
    ($a$): Preferred mass when the progenitor radius from pre-explosion observations was considered. 
    ($b$): From PARSEC stellar models \citep{2012bressan}.
    \label{tab:finalmass}}
    
\end{table*}
%------------------------

%------------------------
\subsection{Preliminary Assumptions}
%------------------------

Many of the methods used to measure progenitor mass rely on an estimate of either pre or post-explosion luminosity which depend on both distance and extinction estimates. 
Additionally, metallicity affects both the mass-loss rates of massive stars and the structure and evolution  of single and binary stars.
In this section we describe the values we adopt for distance, extinction, and metallicity.

%------------------------
%\subsection{Metallicity}
%------------------------
Based on oxygen abundance measurements of three \ion{H}{2} regions close to SN~2017eaw \citep{2013gusev}, \citet{2019vandyk} adopt a slightly subsolar metallicity of $Z=0.009-0.01$ ($[Fe/H]=-0.2$ to $-0.1$) while other authors assume solar metallicity.
When possible, we will use the subsolar metallicity value in the analyses presented in this paper, however, we do not attempt to correct different progenitor analyses for the minor metallicity differences assumed by different authors.

%------------------------
%\subsection{Distance}
%------------------------
The distance to NGC 6946 has been derived many times with many different methods that yield significantly different distances \citep[see] [and references therein for a through discussion]{2019vandyk}. 
However, recent estimates have converged \citep{2018anand,2018murphy,2020goldberg,2019eldridge}.  We adopt a distance of $7.73\pm0.78$ Mpc ($\mu=29.44$) from \citet{2019vandyk} which falls near the middle of the range of these converging measurements.

%------------------------
%\subsection{Extinction}
%------------------------
An accurate estimate of extinction plays a key role in measuring luminosity.
Calculating extinction for SN~2017eaw is not straight forward. NGC~6946 lies near the galactic plane and thus has a relatively high Milky Way extinction. 
Additionally, there may have been more host extinction along the line of sight to SN~2017eaw prior to explosion than is seen after explosion.
This is because RSGs may lose mass in the years preceding explosion which affects pre-explosion observations.
However, the SN explosion itself can destroy this dust, leading to a different post explosion host extinction \citep{2012kochanek2, 2012vandyk, 2012fraser}.

Post-explosion host extinction is often calculated from the equivalent widths of the Na ID and diffuse interstellar band (5870 \AA{}) at the host redshift. 
Unfortunately for this event we are unable to measure the equivalent width of the host absorption features in low resolution spectra because the low recession velocity of NCG 6946 \citep[40 $\mathrm{km\,s^{-1}}$;][]{2008epinat} causes the host dust absorption features to be blended with the strong Milky Way dust features.
Nonetheless, several authors analyze the Milky Way and host extinction simultaneously, arriving at a range of measurements of host extinction: $E(B-V)$=0--0.3 mag \citep{2018kilpatrick,2019rui,2019szalai,2019vandyk}.  The only analysis based on high-resolution spectroscopy finds no evidence for host extinction \citep{2019vandyk}.
Thus, we adopt the Milky Way extinction as E(B-V) = 0.30 mag\citep{2019vandyk} and consider there to be no post-explosion host extinction.

For pre-explosion host extinction, \citet{2019vandyk} find $A_{V}=1.2-4.7$ mag depending on whether the temperature was constrained to typical RSG temperatures.
\citet{2018kilpatrick} find $A_{V}=1.264 ^{+2.6}_{-0.94}$ mag.
Given the consistency of these measurements, we adopt the pre-explosion luminosity and temperatures from \citet{2018kilpatrick}, \citet{2019vandyk}, and \citet{2019rui}, adjusting only the luminosities to the distance of \citet{2019vandyk}.

\subsection{Progenitor Mass Constraints}\label{sec:ProgConstrain}
%------------------------
\subsubsection{Light Curve Constraints}
%------------------------
Progenitor and SN properties can be identified by modeling a SN light curve using a variety of methods from semi-analytic models to radiation hydrodynamic modeling of stellar evolution and the SN explosion.
In these models, care must be taken to differentiate between ZAMS mass and final mass, especially in the case of the analytic models.  It should also be kept in mind that this method is inferring an initial mass from an ejecta mass \citep[which is degenerate with explosion energy, nickel mass, and progenitor radius ][]{Goldberg+2019, 2020goldberg}.
%fundamentally probing the ejecta, nickel mass, explosion energy, and progenitor radius}, and only indirectly inferring the progenitor's initial mass \citep{Goldberg+2019}}. 
We consider the computational models of \citet{2020goldberg} and \citet{2020morozova}.

\citet{2020goldberg} evolve a grid models with Modules for Experiments in Stellar Astrophysics \citep[\texttt{MESA}][]{2011paxton, 2013paxton,2015paxton, paxton_modules_2018, paxton_modules_2019} to iron core-collapse varying progenitor ZAMS mass, surface rotation, mixing length, core overshooting, and wind efficiency.
The SN explosion and subsequent evolution are modelled with \texttt{STELLA} \citep{1998blinnikov,2000blinnikov,2006blinnikov} with a variety of explosion energies.
Using Equation 1 of their paper which relates explosion energy and progenitor mass to luminosity, nickel mass, plateau length, and progenitor radius, they identify a set of $M_{ZAMS}=$14, 19, 22 $\msun{}$ models which are consistent with the observed light curve, nickel mass, and plateau length.
These masses are not exhaustive but are rather demonstrative of the range of masses allowed by the degeneracy of light curve modeling, which allow for a continuous range of masses.
When they include the progenitor radius from \citet{2018kilpatrick}, they find only the model with  $M_{ZAMS}=14$ $\msun{}$ is consistent with the observations of SN~2017eaw at a distance of 7.54 Mpc. 

\citet{2020morozova} model SN~2017eaw with a $M_{ZAMS}=15$ $\msun$ progenitor from the KEPLER code \citep{2016sukhbold}. 
The SN explosion and evolution are modeled with the Supernova Explosion Code \citep[SNEC;][]{2015morozova}.
They find that the 15 $\msun$ model matches the observed light curve and no other progneitor mass was explored.
However, as discussed above, \citet{Goldberg+2019, 2020goldberg} detail the degeneracy between progenitor mass, explosion energy, and progenitor radius in such fitting without additional information such as progenitor radius from pre-explosion observations.
In this case, the relationship between $M_{ZAMS}$ and progenitor radius is fixed by the KEPLER modeling and only degeneracy in explosion energy and mass were explored.

SN light-curve modeling is performed using stellar evolution models from which helium core masses can be drawn.
\citet{2020morozova} use the non-rotating, solar metallicity models of KEPLER models of \citet{2016sukhbold}, that have a helium core mass of 4.4 $\msun$.
\citet{2020goldberg} provide final helium core masses for each of their models in Table 1, giving masses of 4.6, 6.8, and 7.7 $\msun$, for the $M_{ZAMS} = 14,\, 19,$ and $22\, \msun$ models, respectively. 

%------------------------
\subsubsection{Pre-explosion Spectral Energy Distribution Constraints}
%------------------------
Another method that can be used to derive progenitor mass is the direct modeling of the spectral energy distribution (SED) of the progenitor star prior to explosion. 
From pre-explosion images, the luminosity and temperature can be derived and used to find the ZAMS mass of the progenitor star.

The site of SN~2017eaw was observed extensively with both HST and the Spitzer Space Telescope, providing measurements from the ultraviolet through the infrared - an unusually complete SED. 
\citet{2018kilpatrick} derive $\log(L/\lsun)=4.9\pm0.2$ and $\log(T_\star)=3550^{+450}_{-250}$K which they compare to evolutionary tracks to  infer $M_{ZAMS}=13^{+4}_{-2}~\msun{}$. 
We note that when a coarse grid of models is used to determine a progenitor mass, the uncertainty in the ZAMS mass can be dominated by the grid spacing. 
The set of models considered in this analysis were ranged from $M_{ZAMS}$=9--17 $\msun{}$ in increments of 1 $\msun{}$ and thus 13 $\msun{}$ should be considered an approximate value.
We further note that the distance that they use for this analysis $6.72\pm0.15$ Mpc is less than the distance used in this paper. 
A greater distance would increase the luminosity, and the inferred progenitor mass. 
We will account for this fact later in this section. 

\citet{2019vandyk} perform a similar analysis on the progenitor SED, 
They derive $L_{bol}\approx(1.2\pm0.2) \times 10^{5}\lsun{}$ and $T_{eff}\approx2500-3300$ K. 
They find that the bolometric luminosity is only consistent with a progenitor with $M_{ZAMS}=15 ~\msun$ from a grid of of models with initial masses: 12, 15, and 18 $\msun{}$.

\citet{2019rui} use the SED to derive an effective temperature of $T=3550 \pm 100$ K.
A luminosity of $log(L_{bol}/\lsun{}) = 4.88 \pm 0.20$ is determined by applying a bolometric correction \citep{2006levesque} to the K-band magnitude, resulting in $M_{ZAMS}=12 \pm 2 ~\msun{}$. 
Here again a coarse grid of 9, 12, 15, and 20 $\msun{}$ models were used which limit the precision of the progenitor mass derived.

Although RSG luminosity is typically translated to initial mass, it can be used to derive helium core mass at the end of carbon burning \citep[e.g.,][]{2009smartt, 2020farrell}. 
We correct the luminosities derived from \citet{2018kilpatrick} and \citet{2019rui} to the distance of \citet{2019vandyk}, finding $\log(L_{bol}/\lsun{}) = 5.02$ and $\log(L_{bol}/\lsun{}) = 5.17$, respectively.
We then use Equation 1 from \citet{2020farrell} to derive helium core masses of $M_{He, core} = 4.8 ~\msun{}$ \citep{2018kilpatrick}, $M_{He, core} = 5.1 ~\msun{}$ \citep{2019vandyk}, and $M_{He, core} = 6.0 ~\msun{}$ \citep{2019rui}.

%------------------------
\subsubsection{Nebular Spectra Modeling}
%------------------------
An alternate method to measure the final mass of a SN progenitor is based on the strength of the [OI] ($\lambda\lambda 6300,6363$) emission line doublet in spectra taken 200-500 days after explosion \citep{2014jerkstrand, 2020dessart}. 
At this time the outer layers of the SN ejecta are transparent, revealing the inner nucleosyntheic products created during the evolution of the star.
In particular, the oxygen mass, which can be derived from the oxygen luminosity, is strongly related to the helium core mass which itself relates to the ZAMS mass \citep{1995woosley}.

One way to derive progenitor mass from nebular spectra is to model the spectroscopic evolution of the [OI] emission doublet starting from the explosion of an evolved RSG.
Using spectra taken 220, 435, and 491 days after explosion, \citet{2019szalai}
 find the oxygen luminosity of SN~2017eaw is best matched by the $\mathrm{M_{ZAMS}=15} ~\msun{}$ model of \citet{2014jerkstrand}.
Although \citet{2019szalai} uses a different distance and extinction, the empirical scaling accounts for these differences and no modification to the derived mass is needed for this paper.
\citet{2019vandyk} compare a different set of late-time spectra (taken on days 213.9, 245.9, and 415.2) of SN~2017eaw to the same models \citep{2014jerkstrand} and also find them to be most consistent with an $\mathrm{M_{ZAMS}=15} ~\msun{}$ progenitor. 
In both cases, these masses are derived from a grid of models with initial masses: 12, 15, 19, and 20 $\msun{}$.

The nebular spectra of SN~2017eaw are evaluated using the models of \citet{2014jerkstrand}.
These models use the non-rotating solar metallicity KEPLER models from \citet{2007woosley} as input. 
and are included in \citet{2016sukhbold} where the helium core masses are tabulated in Table 2. 
The 15 $\msun$ model has a final helium core mass of 4.35 $\msun$ which we use as the final helium core mass derived from nebular spectra.

In summary, when considering all of the available constraints on the final state of the progenitor and the spectroscopy of the event itself, the most likely He core mass of the SN~2017eaw progenitor is 4.4--6.0 $\msun{}$ ($M_{ZAMS} = $13.4--17.6 $\msun{}$). 
%We adopt these values to determine He core mass in \autoref{sec:final_he_core} which enables us to compare population ages to these progenitor masses in \autoref{sec:discussion}.

%------------------------
\subsubsection{Stellar Populations Constraints}\label{sec:popConstrain}
%------------------------
Over the past decade, there have been many successful studies of SN progenitor ages evaluated through studies of the star formation history (SFH) of the very nearby surrounding resolved stars \citep{badenes2009,gogarten2009,jennings2012,2014williams}.  Here we discuss previous attempts by \citet{2018williams} and \citet{2021koplitz} to apply this technique to SNe in NGC~6946.
\citet{2018williams} measured a possible spike in star formation around 30 Myr, corresponding to an initial mass of $8.8^{+2.0}_{-0.2}$ $\msun$ and He core mass of $\sim$2.7 $\msun$ using PARSEC stellar isochrones \citep{2012bressan}.
\citet{2021koplitz}, meanwhile, did not find evidence for any significant young population, although their upper limit was consistent with the other result at that age.

There were a few key differences in the analyses of \citet{2021koplitz, 2018williams}.
First, they differ in whether or not a widespread background stellar populations was simultaneously fit when deriving an age from the SFH.
Simultaneously fitting the background populations while fitting the SFH adds more weight to populations unique to the SN region.
In \citet{2021koplitz}, the background stellar population is modeled as a contaminating population consisting of the stars in an annulus from just outside the SN region (in this case 50 pc) to 1000 pc, and is scaled to the size of the SN region before fitting. 
\citet{2018williams} also adopted a closer distance (6.8 Mpc), which could lead to an overestimate of the age of any detected young stars.
Finally, the selection criteria for which stars were included in the analysis were not the same between the analyses.
We discuss our reanalysis optimized for SN~2017eaw in \autoref{sec:reanalysis}.

\section{Inclusion of binary scenarios for Type II SN progenitors}\label{sec:modeling}

%The motivation for this paper is to understand the discrepancy in the literature between the inferred initial mass, as derived from the age of the stellar population surrounding SN~2017eaw compared with the ones derived from a variety of pre- and post-explosion observational techniques. As we summarize in \autoref{tab:finalmass}, the methods that probe the final progenitor state infer an initial mass in the bulk range of $12-15  \msun$, whereas previous works, using the age of the population, found an initial mass of $\leq 10.8\msun$.

\citet{Zapartas+2019,Zapartas+2021} studied the expected age of the Type II SN progenitors, employing single and binary population synthesis models. They find that the delay-times between the birth of a binary progenitors and its eventual core-collapse is on average higher than single star progenitors. In fact, binary progenitors can have delay-times that surpass the maximum age threshold of single star progenitors of around $50$ Myrs, reaching to ages of 200 Myrs \citep{Zapartas+2017a}. The long delay-time is predominantly due to the low initial mass of the SN progenitor. Even though it eventually evolves to a higher mass via binary evolution, these low mass progenitors have longer evolutionary timescales, which dictate the evolutionary timescale of the whole binary system.

The main binary scenarios that lead to Type II SNe involve the increase of mass of the SN progenitor, during either a stable mass accretion phase or during a faster merging process with its companion  \citep[e.g.,][]{Podsiadlowski+1992,Zapartas+2019}. 
In both cases, the final binary product has a higher total and more importantly helium core mass at the end of its life, compared to the one it would form without any mass gain from its binary companion. The increase of the core mass of these binary Type II progenitors is expected to lead to an increase in their preSN luminosity. Thus, observational methods that directly probe the preSN core mass or luminosity would infer a higher initial mass for the progenitor if they consider only a single star evolution. However, as mentioned above, these binary progenitors would have longer delay-times compared to a single star of the same initial mass (as inferred from the final mass). 

Based on these findings, \citet{Zapartas+2021} suggested that a discrepancy between the inferred initial mass of the progenitor from observational methods that directly probe the progenitor's core in its preSN state and the age estimates of its surrounding population could be an indication of a binary progenitor\footnote{Note that in their analysis the scenarios for Type II SN progenitors from partially stripped donor stars was not found as a dominant one, although it has been investigated in works employing detailed stellar models \citep{Eldridge+2018,Eldridge+2019}. Still, the latter path is not expected to affect the total lifetime of the progenitor, so it would not be able to explain a inferred mass discrepancy as discussed here.}.

In \autoref{sec:discussion} we consider both single and binary evolution channels to interpret our observational results. For this, we use the same set of binary population synthesis models of core-collapse SNe from single and binary progenitors as in \citet{Zapartas+2019, Zapartas+2021}. These models employ the binary population synthesis code {\tt binary\_c} \citep{Izzard+2004,Izzard+2006,Izzard+2009}, based on \citet{Hurley+2000} fitting formulae for single stars and taking into account physical processes in binary systems \citep[][including metallicity of $Z=0.008$, very close to the value of the SN~2017eaw region, no rotation, convective overshooting according to \citet{Pols+1998}, and wind mass-loss prescriptions throughout the evolution as described in \citealt{Zapartas+2017a}]{Hurley+2002}.

For our population synthesis models with the closest metallicity to SN2017eaw site (Z=$0.008$) we recover a relation to derive single-star initial masses from final helium core masses of 

\begin{equation} \label{eq:helium_zams_singles}
\frac{M_{\rm He-core,final}}{\msun} = 0.34 \frac{M_{\rm initial}}{\msun}+ 0.002 \big(\frac{M_{\rm initial}}{\msun}\big)^2 - 0.56,
\end{equation}
very similar to Eq.1 of \citet{Zapartas+2021} for solar metallicity. 

%%%%%%%%%%%%%%%%%%%%%%%%%%%%%%%%%%%%%%%%%%
%\section{Analysis of 2017eaw progenitor}\label{sec:analysis}
%%%%%%%%%%%%%%%%%%%%%%%%%%%%%%%%%%%%%%%%%%

%In addition to taking values from the literature concerning the progenitor mass of SN~2017eaw, we derive a new progenitor age and helium core mass, which we will use to further comment on the potential binary nature of the progenitor. We discuss each of these measurements below and list them in \autoref{tab:finalmass}.

%------------------------
\section{Reanalysis of the Surrounding Population}\label{sec:reanalysis}
%------------------------
With the broad agreement of final progenitor mass from a variety of methods and observations, we reexamine the age of the SN region to uncover the source of the discrepancy. 
Given that both \citet{2018williams} and \citet{2021koplitz} included SN~2017eaw as part of a sample analysis where a uniform analysis technique was applied to all SN sites in the sample, we reanalyzed this region, optimizing our technique for the specific characteristics of this progenitor location.
The results of our new age analysis are summarized in \autoref{tab:pop_fits}.
Details of the fitting technique are provided below. 

To update the color magnitude diagram (CMD) fitting results, we used the package MATCH \citep{2002dolphin,2012dolphin,2013dolphin} which attempts to reproduce an observed CMD using PARSEC stellar isochrones \citep{2012bressan}.
MATCH returns the best-fit star formation rate as a function of look-back time as well as the upper and lower uncertainties on this rate.
The SN progenitor is then assumed to belong to the median population younger than a chosen age limit.  The full details of the fitting and age-inference technique are described in \citet{2021koplitz}.  Here we adjust the area on the sky from which we take our sample, the photometric cuts applied to the catalog, the age limits of the populations considered, and the inclusion or exclusion of a background component in the fitting.

\begin{figure}[ht]
\begin{center}
 \includegraphics[width=3.6in]{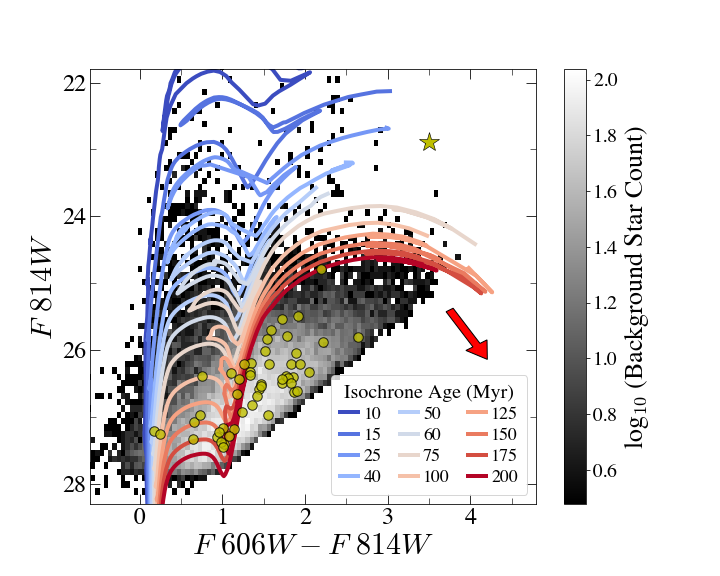}
 \caption{CMD of the stars within 50 pc of SN~2017eaw shown as yellow circles with the background populations shown in gray scale. Overplotted are isochrones between 10 and 200 Myr with blue (red) representing younger (older) ages. The progenitor star is highlighted in the top right corner of the figure and is shown as a star. The red arrow shows the impact of reddening on our photometry.}
 \label{fig:cmd}
\end{center}
\end{figure}

Additionally, while \citet{2021koplitz} focused on stellar populations younger than 50 Myr, binary evolution can produce Type II SNe from stellar populations as old as 200 Myr \citep{Zapartas+2017a}. 
Unfortunately, comparing model isochrones to our CMD in \autoref{fig:cmd}, the sensitivity of our photometry limits our analysis to 150 Myr rather than 200 Myr. 
Still, the expected SN rate from all binary scenarios with delay-times of 150-200 Myr is of the order of $3 \times 10^{-6}$ Myr$^{-1}$ $M_\odot^{-1}$ (see eq. A2 of \citealt{Zapartas+2017a}), consisting around 1\% of all SNe.
Thus, we encompass the vast majority of all possible single and binary progenitor scenarios within the 150 Myr limit from photometry. 

Note that in our analysis, we only take into account the possibility of a binary interaction of the  progenitor of SN~2017eaw, not considering the effect of binary products on the surrounding stellar population when we infer the age of the population (something independent of the origin of the progenitor itself). Binary products are expected to make a host region of a Type II SN look slightly bluer, and thus ``younger'' \citep{Xiao+2019, Schady+2019}.

%------------------------
%------------------------

We start with the photometry catalog of \citet{2021koplitz}; however, we correct the astrometry since there was a slight misalignment in their analysis.
The offset was small enough that many of the same stellar populations were included in both their fit and ours.
Additionally, we loosened the quality cuts to signal-to-noise ratios above 3.5 and sharpness squared values below 0.5, which were designed to produce the best photometry for the crowded regions of the disk. 
Since SN~2017eaw resides in the outer disk of NGC~6946, DOLPHOT is able to make good measurements at its location resulting in the low crowding values.
In fact, the crowding cut was found to be unnecessary as every object that passed our signal-to-noise and sharpness cuts had crowding values less than 1 (i.e. good stars). Our new catalog includes more stars and increases our sensitivity to any young population.

MATCH finds total extinction (MW + host) by applying three types of reddening: A$_V$, a universal extinction for all stars, dA$_V$, a differential extinction calculated for each star individually, and A$_{VY}$, an additional differential extinction applied to stars younger than 100 Myr old.
The best-fit A$_V$ and dA$_V$ provide the range of possible total extinctions for the region, i.e. between A$_V$ and A$_V$ + dA$_V$ (+A$_{VY}$ for young populations).
The details of the different types of extinction used by MATCH can be found in \citet{2021koplitz}.
Most of our fits found an A$_{V}$ value of $\sim$1.4 with little dA$_V$ ($<$0.4).
This is slightly higher than the Milky Way contribution toward NGC 6946 ($A_{V}=0.938$; \citealt{2011schlafly}), but is consistent with little to no host contribution, similar to what other techniques have found \citep{2018kilpatrick,2019vandyk}.

To understand how nearby objects affected the our age estimates, we ran MATCH on a variety of photometric catalogs derived using different extraction radii.
In this way we account for the possibility that the progenitor of SN~2017eaw traveled away from its birth place as a result of an ejection from a binary system. 
These runaway progenitors can travel typically $\sim$ 100 pc until their own explosions \citep{Renzo+2019}.
We thus conservatively considered radii ranging from 50 to 200 pc in steps of 50 pc to encompass the population at the birth site of a runaway progenitor. 
\autoref{fig:region_summary} shows the location of SN~2017eaw, indicating the 50, 100, and 200 pc radii used in our analysis with white circles. 
A nearby ($\sim$200 pc) young star cluster is shown with a red circle.
In \autoref{sec:runaway} we argue that this cluster is unlikely to be the origin of the progenitor of SN~2017eaw.
As it dominated our fits with radii $\geq$150 pc, we did not consider the 200 pc radius result for the remainder of our analysis. 

Additionally, we fit the SFH considering all stars near the SN site equally as well as down weighting those consistent with a background populations.
When MATCH simultaneously fits the background population, it models and down weights a background population before determining the SFH of the remaining population.
Not fitting a background component allows us to determine how much overall SF is present while the fits with the background allow us to isolate SF unique to the region.

Finally, we consider the effects of the inclusion of the progenitor of SN~2017eaw in our final catalog.
To determine how the progenitor impacts our inferred ages, we fit the CMD of each radius without it. 
When we fit a separate background population simultaneously, the median age did not change from the fit with the progenitor. 
In the fits without the background populations, the median age did not change whether or not the progenitor was included; however, the uncertainties were smaller without the progenitor. 

%------------------------
\begin{table*}[ht!]
    \centering
    \begin{tabular}{|c|c|c|c|c|c|}
    \toprule
    Radius & Num. of & 50 Myr & 50 Myr & 150 Myr & 150 Myr\\
    $\mathrm{(pc)}$ & Stars & Background & No Background & Background & No Background\\
    \hline
    50 & 50 & --- & $16.8^{+3.2}_{-1.0}$ & --- & $85.9^{+3.2}_{-6.5}$\\
    100 & 210 & --- & $16.6^{+3.4}_{-4.0}$ & $84.1^{+5.0}_{-27.9}$ & $148.8^{+9.7}_{-7.5}$\\
    150 & 558 & $15.0^{+5.0}_{-2.4}$ & $22.9^{+2.2}_{-10.3}$ & $93.0^{+7.0}_{-13.6}$ & $142.0^{+16.5}_{-29.8}$\\
    \botrule
    \end{tabular}
    
    \caption{Summary of our stellar population fitting. Column (1) shows the radius of the region being considered, centered SN~2017eaw, in units of pc. Column (2) is the number of stars used in the fit. Columns (3) and (5) are the median age, in units of Myr, when we simultaneously fit the SN and background populations with a 50 and 150 Myr cut off, respectively. Fits with no SF at ages younger the cutoff are shown as dashed lines. Columns (4) and (6) are the median age when the background population is not considered a separate population with a 50 and 150 Myr cut off, respectively. 
    %Column (7) shows any possible contaminating sources in the fit.
    \label{tab:pop_fits}
    }
\end{table*}
%------------------------

%------------------------
\begin{figure*}
    \centering
    \includegraphics[width=0.85\textwidth]{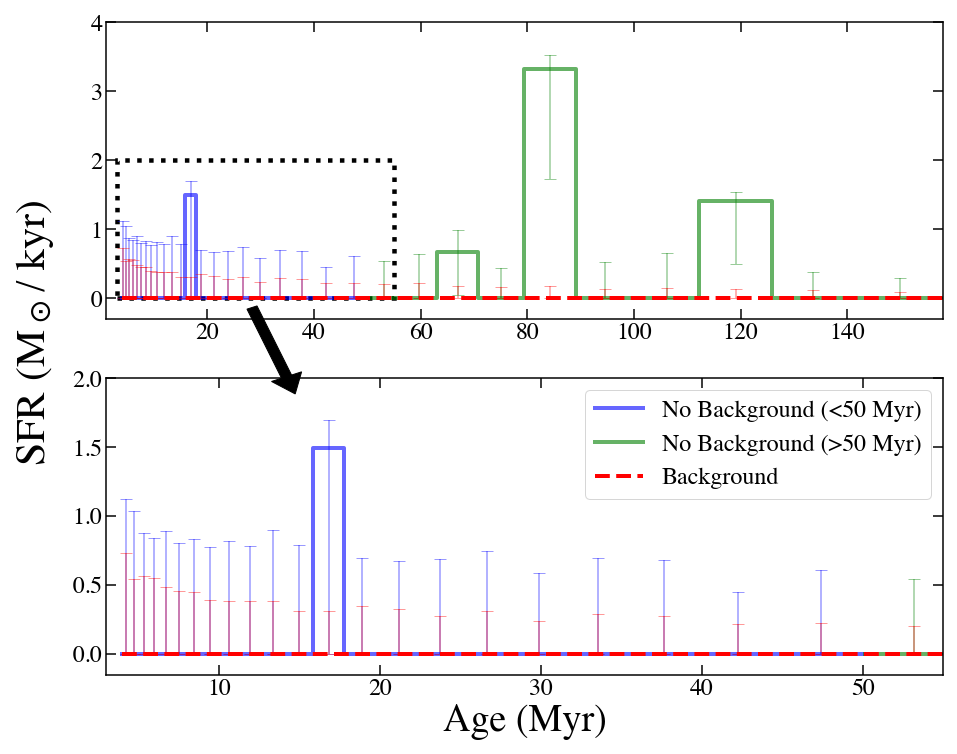}
    \caption{SFHs measured with a 50 pc exaction radius. The red lines show the SFH when the background populations were simultaneously fit. The fit without the background populations are shown in blue for those younger than 50 Myr and green for those older. The top panel shows the most recent 150 Myr of the SFHs and the bottom panel zooms in on the first 55 Myr, highlighting the young populations measured. The black dotted box in the top panel indicates the limits of the bottom panel plot.}
    \label{fig:sfh_150myr}
\end{figure*}
%------------------------

In addition to fitting multiple radii and with and without a background population, we consider both a young stellar population only (upper age limit of 50 Myrs, consistent with limits single star evolution) and an additional older stellar population (upper age limit of 150 Myrs, consistent with limits from binary evolution). Our fits with a 50 pc extraction radius and including a background CMD found no populations younger than 150 Myr. 
Not finding young SF above the background level near the SN site is not surprising given that the location is extremely sparse while NGC~6946 has a high SF rate, between 3.2 and 12.1 M$_\odot$yr$^{-1}$ \citep{2013jarrett, 2019eldridge}.
Given that, had this analysis been applied to a SN remnant, with no knowledge of SN type, \citet{2021koplitz} likely would have concluded that it was a Type Ia SN. 
Care should be given to the use of background populations in spare regions and highly SF galaxies. 

Fitting without the background CMD shows that there are young populations near SN~2017eaw, with the 50 Myr cut off finding a median population consistent with progenitor age estimates from near-explosion modeling.

Even with a 100 pc radius, our fits with a background component continue to find no populations younger than 50 Myr, though a burst in SF is seen at 84 Myr when we expand the ages considered to 150 Myrs. 
With a 50 Myr cut off, our fit without a background component measures the same median age for both a 50 and 100 pc radius.
Thus, this age component of the population appears to be widespread, partially explaining why it is considered a background population.
At larger radii and older  ages (150 Myr), when there is no background component included, a 150 Myr old population always dominates the fit.

Our fits including a background component only detect SF younger than 50 Myr with a 150 pc radius, though it is a small amount ($\sim$1 M$_\odot$ kyr$^{-1}$).
The median age ($15.0^{+5.0}_{-2.4}$ Myr, corresponding to an He core mass of $\sim$$4.5 - 4.7$ $\msun$; \citealt{2012bressan}) is consistent with our estimate with smaller radii without a background population and those from near-explosion.  These are therefore the same population, but it appears to be spread out over a large enough area that it is hard to detect as separate from the background at smaller radii.

Our progenitor age remains consistent when we only consider the stars within 150 pc, though the median is older.
It is not surprising that the median age is slightly older when no background component is included.
\citet{2023koplitz} showed that progenitor ages from fits without a background component were older than fits with one, likely because the background population is more likely to be comprised of older stars that have had more time to disperse.

%------------------------
%\subsection{UV Age Estimate}
%------------------------
In addition to the optical data presented above, \citet{2023Tran} derived a spatially resolved SFH map of NGC 6946 from HST/WFC3  stellar photometry in the NUV bands F275W and F336W for ages younger than $\sim$30 Myr.
They used a quadtree algorithm to divide their NUV ($F275W$ and $F336W$) HST photometry into grids based on stellar density, with the smallest grids in regions of highest stellar density.  
SN~2017eaw lies primarily in one large grid region of 647 square arcseconds ($\sim$ 937 pc x 937 pc) since it is located in an area of low stellar density.
The median age of the young ($\lesssim$30 Myr) populations in that region was found to be $12.6^{+20.0}_{-12.6}$ Myr, consistent with the 4.4--6.0 $\msun{}$ He core mass ($M_{ZAMS} = $13.4--17.6 $\msun{}$) %$\sim$15 $\msun$ progenitor mass 
found from near-explosion observations and our population age derived from optical observation. Due to the multipeaked nature of the SFH used to derive the median age of this population, the spread in the uncertainty, which is derived from bootstrapping the SFH, obtained is very high. In addition, the youngest time bin of the star formation history goes from 0 to 4 Myr, which contributes to the large lower limit on uncertainty.

%------------------------
\begin{figure}[ht]
\begin{center}
 \includegraphics[width=3in]{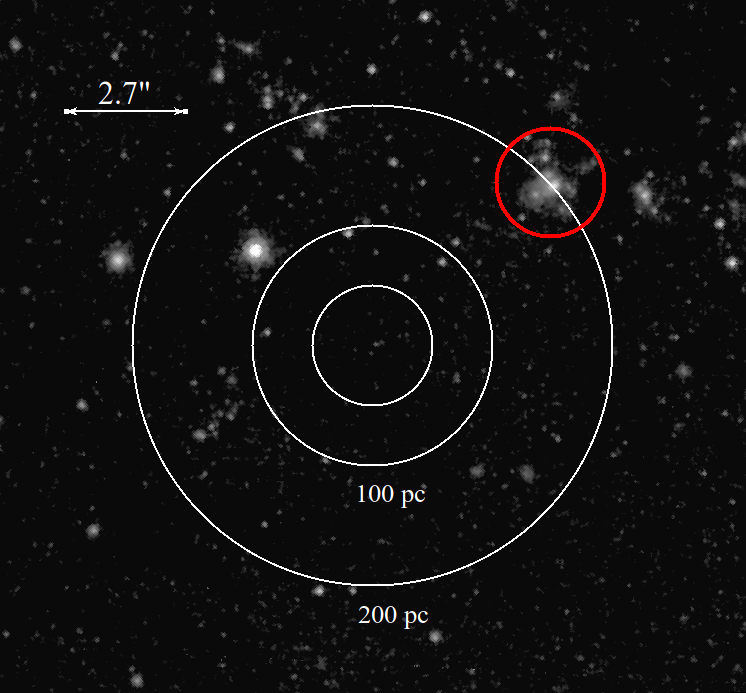}
 \caption{Our $F606W$ image of the region around SN~2017eaw. The white rings indicate our extraction regions with 50, 100, and 200 pc radii. The red circle indicates the young stellar cluster that seems to dominate our fits beyond 150 pc.}
 \label{fig:region_summary}
\end{center}
\end{figure}
%------------------------

Given that when the background is considered, no star formation is found within 50 pc of SN~2017eaw when either maximum age is used and within 100 pc of SN~2017eaw when the 50 Myr cut off is used, we find it necessary to consider that SN~2017eaw may have come from the background population, which does appear to contain a young component, as found in both the \citet{2023Tran} map and our new analysis here. 
Thus, the most likely progenitor age is 16.8 Myr (corresponding to a $M_{ZAMS} = 14$ $\msun$ and a He core mass of $\sim$$3.7 - 4.7$ $\msun$; \citealt{2012bressan}) from our fit to the 50 pc region with no background component.

While it is tempting to consider the progenitor of SN~2017eaw a single star and declare the discrepancy solved with the population found at 50 pc when no background is considered using a 50 Myr cut-off,
it is still possible that the progenitor of SN~2017eaw was a binary belonging either to the younger population found with the 50 Myr cutoff or the older population found within the same radius but with a 150 Myr cut off.
Next, we will examine the constraints placed by these ages.

%------------------------
\begin{figure*}[ht]
\begin{center}
 \includegraphics[width=4.9in]{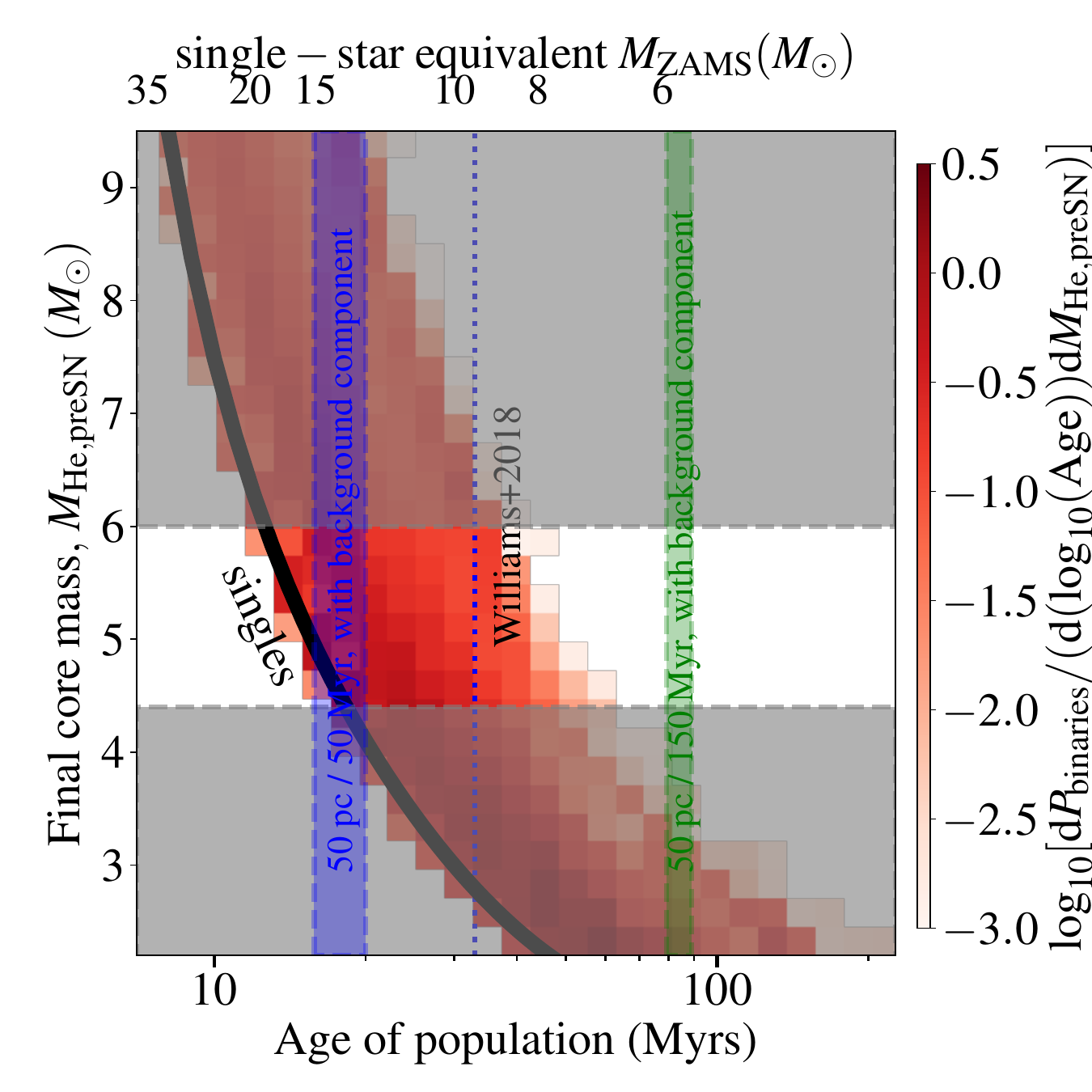}
 \caption{Correlation of preSN helium core mass with the age of a Type II progenitor systems. Black line is the unique correlation for our single star models at $Z=0.008$. %$Z=Z_{\odot}$. 
 The 2D probability density of binary progenitors (red density map) spreads to higher ages compared to single stars (for the same final helium core mass). For SN~2017eaw, we show with grey the constraints between $4.4 - 6.0 ~\msun$ helium core mass from prior preSN imaging, light curve modelling, and nebular spectra. 
 The two possible population ages found in this work for the 50 pc region around SN~2017eaw,  $16.8^{+3.2}_{-1.0}$ ($85.9^{+3.2}_{-6.5}$) Myrs, allowing for star formation up to 50 (150) Myrs with no simultaneous fitting of a separate background component, are shown with blue (green) vertical shaded regions (first line of \autoref{tab:pop_fits}).
 The dotted, black line is the prior age estimate for 2017eaw from \citet{2018williams} of $\sim 33$ Myrs without error bars.
\label{fig:2D_age_Hecore} }
  
\end{center}
\end{figure*}
%------------------------

%%%%%%%%%%%%%%%%%%%%%%%%%%%%%%%%%%%%%%%%%%
\section{Possible progenitors}\label{sec:discussion}
%%%%%%%%%%%%%%%%%%%%%%%%%%%%%%%%%%%%%%%%%%
We consider the following scenarios for the progenitor of SN~2017eaw: an isolated initially single star, an interacting binary system, and a runaway star.
%------------------------
\subsection{The single or binary progenitor scenario}
%------------------------
\autoref{fig:2D_age_Hecore} summarizes the results of our binary population synthesis and can be used as a theoretical framework for our finding with the new analysis of SN~2017eaw. 
There is a unique anti-correlation between the preSN helium core masses of single stars and their lifetimes prior to core collapse (solid black line) which is given by \autoref{eq:helium_time_singles}.

%123.71487754 -298.69026391  289.25883316 -128.46234239   21.71348526
%\begin{eqnarray} \nonumber \label{eq:helium_time_singles}
\begin{align} \nonumber \label{eq:helium_time_singles}
&\frac{M_{\rm He-core,final}}{\msun} = 123.71 -  298.69  \log_{10}\left(\frac{t}{\rm{Myr}}\right) + \\ 
&+ 289.26 \left[\log_{10}\left(\frac{t}{\rm{Myr}}\right)\right]^2 - \,\, \\ \nonumber
& -128.46 \left[\log_{10}\left(\frac{t}{\rm{Myr}}\right)\right]^3 + 21.71 \left[\log_{10}\left(\frac{t}{\rm{Myr}}\right)\right]^4,
\end{align} 
%\end{eqnarray} 
valid for ages between roughly 8-50 Myrs (typical Type II single star progenitor lifetimes).  Single star progenitors of higher core masses are expected to have shorter lifetimes. 
This relation is equivalent to our \autoref{eq:helium_zams_singles} for initial masses, but again only for single star progenitors of Type II SN.

For binary progenitors the same general trend of shorter lifetimes for higher core masses can be seen  too (red 2D density distribution). However, the different evolutionary scenarios break the unique correlation, leading to a spread towards higher delay-times between the birth of the original stellar system and the SN event itself.

%------------------------
\begin{figure}[ht]
\begin{center}
 \includegraphics[width=3.6in]{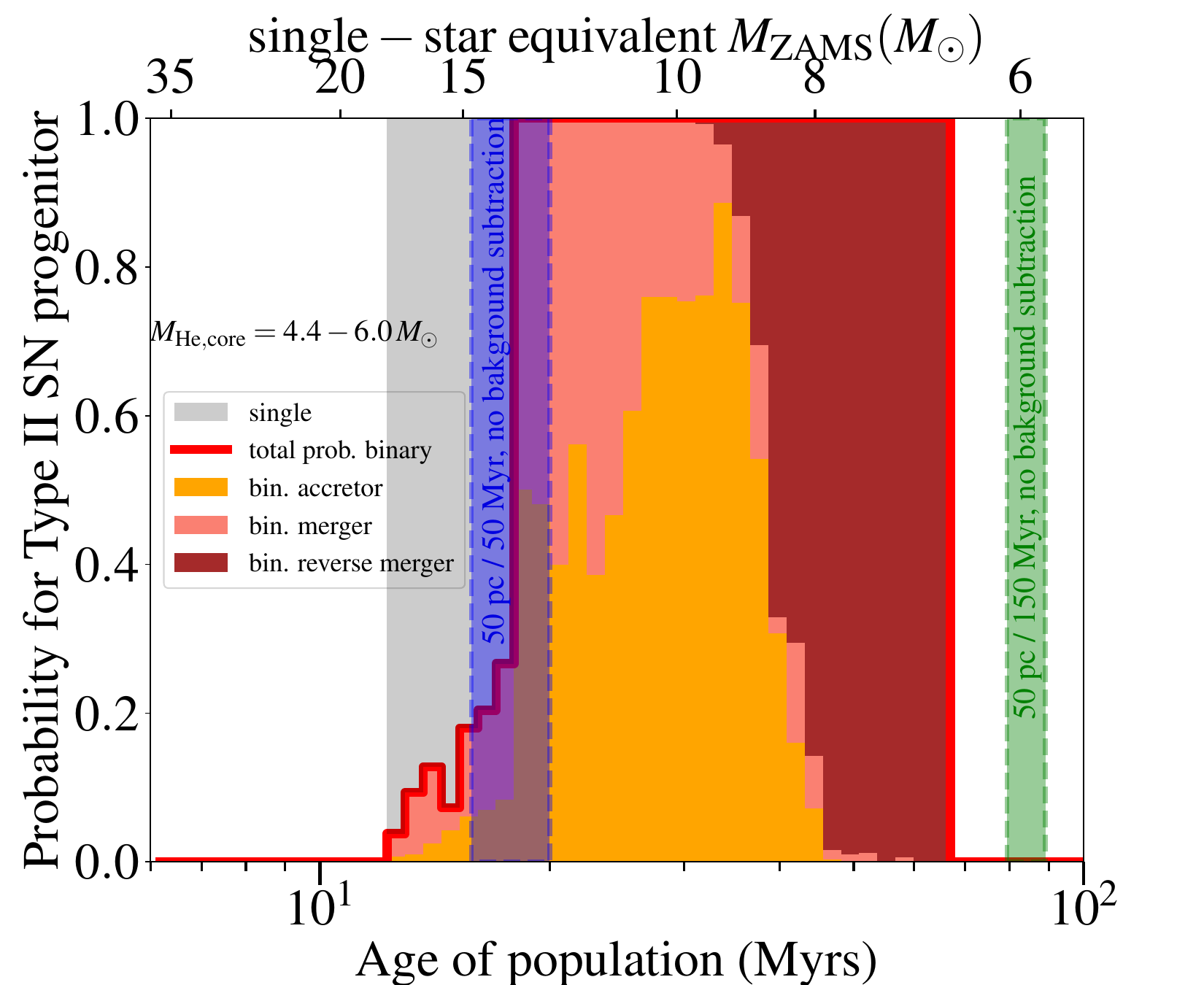}
 \caption{Probability of each evolutionary scenario as a function of measured population age, for progenitors of Type II SNe with final helium cores between $M_{\rm He,core} = 4.4 - 6.0 \, M_{\odot}$, corresponding to the inferred core mass range of SN~2017eaw progenitor. The total probability of all binary scenarios (red line) is subdivided into mass gainers (yellow), MS mergers (pink) and reverse mergers (brown). The gray shaded region shows the probability of a single star progenitors, as a complement of binary probability, and is concentrated to ages between $11-18.2$ Myrs. In white regions, no scenario can reproduce a progenitor with that age. The two possible population ages found in this work for the 50 pc region around SN~2017eaw,  $16.8^{+3.2}_{-1.0}$ ($85.9^{+3.2}_{-6.5}$) Myrs, allowing for star formation up to 50 (150) Myrs with no simultaneous fitting of a separate background component, are shown with blue (green) vertical shaded regions (first line of \autoref{tab:pop_fits}). The former, blue, falling on the high side of the single-star age regime and the latter, green, is inconsistent even with the long delay-time of binary scenarios.
 \label{fig:prob_progenitor_typeII}}
  
\end{center}
\end{figure}
%------------------------

From studies of preSN images, light curve modeling, and late nebular spectra we inferred a helium core mass between $4.4-6.0 ~\msun$  for the SN~2017eaw progenitor (\autoref{sec:final_he_core}; \autoref{tab:finalmass}). 
Combining these values with our binary population synthesis results, the previous age estimate of SN~2017eaw from \citet{2018williams} of $\sim 33$ Myrs clearly falls outside the tight single star relation and very comfortably in the range of binary progenitors. 

In this work, we have found more robust age estimates, with a burst of star formation at either $16.8^{+3.2}_{-1.0}$ (when a 50 Myr maximum age is applied; blue shaded vertical regions) or at $85.9^{+3.2}_{-6.5}$ Myrs (when a 150 Myr maximum age is applied; green shaded vertical regions). Here we see that combining the final mass constraints with the low age derived from 50 Myr cutoff, the progenitor is consistent with both single and binary scenarios.
However, the longer age derived from 150 Myr cutoff combined with the constraints from the final mass is inconsistent with both the single and binary scenarios.

To further quantify these results, in \autoref{fig:prob_progenitor_typeII} we show the probability of different evolutionary scenarios of forming a Type II SN progenitor with a $4.4-6.0 ~\msun$ final helium core, as a function of their delay time from birth. Single star models that form a final helium core in the range above have a delay time between $\sim 18.2$ and $11$ Myrs, respectively (gray area).  

\begin{figure}[ht]
\begin{center}
\includegraphics[width=3.6in]{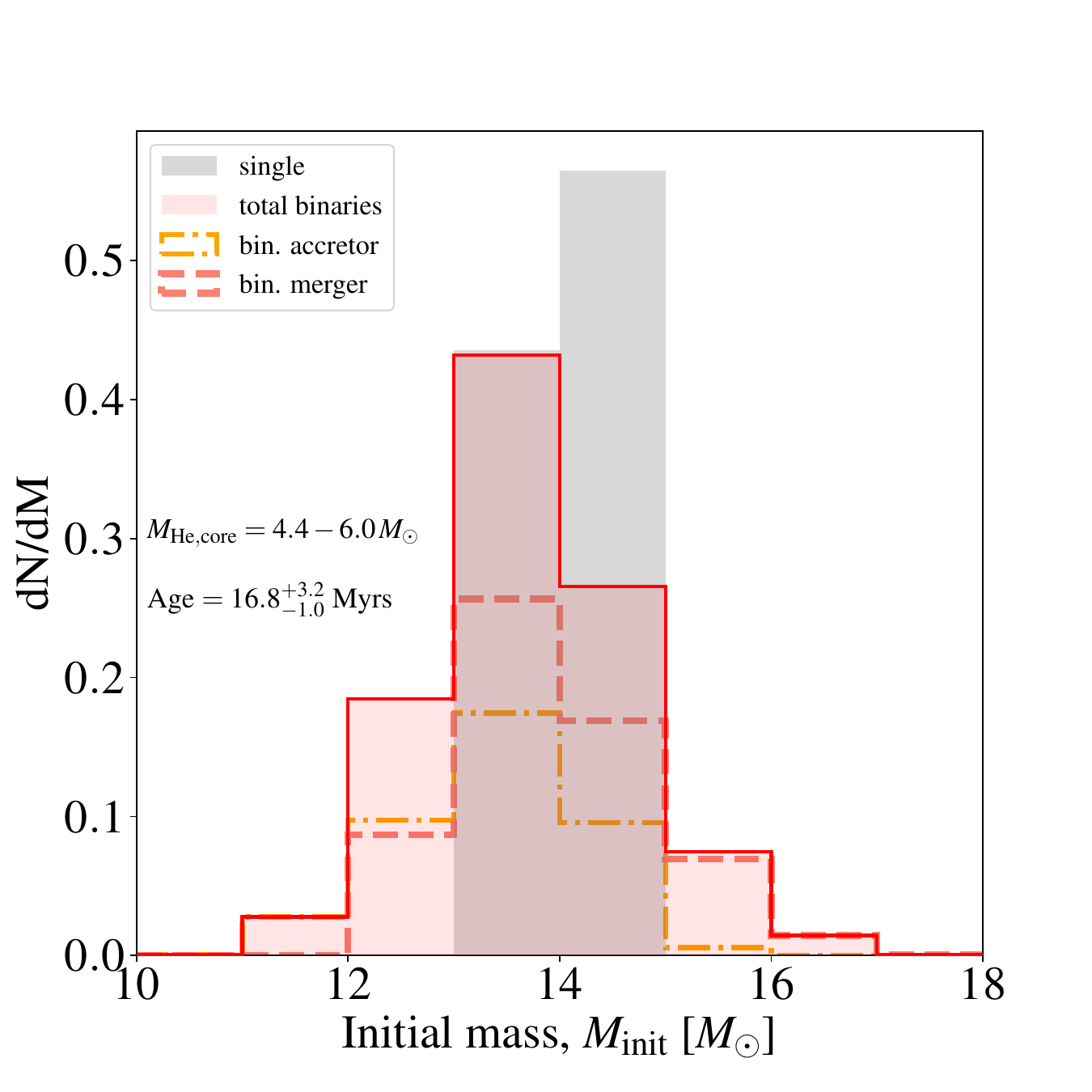}\\
 \caption{Initial mass distribution of Type II SN progenitors, with final helium core, $M_{\rm He,core} = 4.4 - 6.0 ~\msun$ 
 and an age estimate between $16.8^{+3.2}_{-1.0}$ Myrs, similar to SN~2017eaw-like events. The normalized distribution of all binaries (red solid line) is subdivided into binary scenarios (mass gainers or MS mergers), and includes both lower and higher mass progenitors than the normalized distribution of single star progenitors.
  \label{fig:initial_mass}}
 
\end{center}
\end{figure}

According to our models we find around 65\% of progenitors with observationally inferred delay-times between $16.8^{+3.2}_{-1.0}$ Myrs originate from single stars, although this age range is at the older end of their $11-18.2$ Myrs range mentioned above. 
%Our observationally inferred delay-time of the progenitor between $16.8^{+3.2}_{-1.0}$ Myrs predominantly points towards a single star scenario for the progenitor of SN~2017eaw, although on its high side. 
%We find a $\sim 61\%$ probability for the progenitor of SN2017eaw to be a single star according to our models. 
This likelihood is based on the fact that the majority of Type II SNe are expected to originate from single star progenitors \citep[][the latter based on the same set of calculations as in this study]{Podsiadlowski+1992,Zapartas+2019}. The $16.8^{+3.2}_{-1.0}$ Myr age range corresponds to the lifetime of single stars that form  helium core masses between $ 4.0 - 4.9 M_{\odot}$ (using \autoref{eq:helium_time_singles}) and initial masses of $\sim 13-15 ~\msun$ (\autoref{fig:initial_mass}, grey).
%, which partially overlaps with the $ 4.4 - 6.0 M_{\odot}$ extra constraint on the core mass. 
%These single stars have an initial mass of $\sim 13-15 \msun$,}  (\autoref{fig:initial_mass}, grey), 
These masses are very consistent the inferred masses of previous studies \citep{2019vandyk, 2018kilpatrick, 2019szalai, 2020morozova, 2020goldberg, 2019rui} for SN~2017eaw 
and the helium core masses we derive from these studies.
In fact, for single stars, the age of the population constrains the helium core mass to the a small range at the low end of the values inferred from those studies, and could be used to break some of the degeneracy in light-curve modeling even when no pre-explosion imaging exists. 

Interestingly, even with an age estimate in this range, a binary history for the progenitor cannot be excluded, still being a significant possibility with $\sim 35\%$ of the scenarios involving binary interaction (red line, combination of all binary scenarios). Progenitors of these channels also form a helium core mass 
%(the physical property that we argue is more tightly constrained from contemporary-to-SN observations) 
of $4.0-4.9 ~\msun$, $16.8^{+3.2}_{-1.0}$ Myr after birth. However, in these cases, due to prior mass accretion or merging, the star that eventually explodes originates from a broader initial mass range of $\sim 11-16 ~\msun$ (\autoref{fig:initial_mass}, red bars), having a peak of the distribution close to the equivalent single star.

These binary scenarios include the accretion of mass onto the Type II progenitor from an initially more massive binary companion that has already exploded and ejected from the system (yellow), or, slightly more likely, the merging with a companion (pink). 
In the case of mass accretors and in $\sim 1/3$ of the cases of mergers,  the progenitor star gains mass while still on its MS, becoming  $\sim 15~\msun $. For the remainder of the merger cases, the progenitor has already evolved towards becoming a giant before triggering a common envelope phase and leads to the merging with a main sequence (MS) companion. 
In all cases, the binary product is expected to continue its evolution in isolation until death, leading to a final helium core within the constraint and with a total delay-time only slightly higher than that of a single star of the same mass. 

%------------------------
%------------------------

For ages above $\sim 18.2$ Myrs, the probability of a single star progenitor for SN~2017eaw drops to zero. This is because, according to our models, a single star massive enough to produce a final $M_{\rm He,core} = 4.4-6.0 ~\msun$ cannot exceed this delay-time. Without adopting different physical assumptions (e.g. high rotation velocity, higher convective overshooting and/or lower metallicity), longer delay-times can only be reached from progenitors that gained mass or merged with a binary companion during their evolution.

Progenitors that accrete mass through stable binary mass transfer (yellow distribution) result in a peak of the probability of this scenarios at around 30 Myrs. Scenarios involving the merging of the two MS and/or giant stars (pink) have a similar distribution of ages. 
Higher ages of Type II SNe would be reached from progenitors experiencing a reverse merger. These involve the merging of a stripped degenerate carbon-oxygen or oxygen-neon-magnesium core which is engulfed by the initial secondary star when the latter evolves to become a giant. The initial mass of one or both stars is below the canonical minimum threshold for a core-collapse SN, of $\sim 8 ~\msun$. 
However, even this scenario struggles to reach a delay-time as high as $85.9^{+3.2}_{-6.5}$ Myrs that we find when using the 150 Myr limit for maximum age, although some extremely low mass reverse mergers can fall within the uncertainty range, with delay-times of $\gtrsim 70$  Myrs. 
This means that delay times of longer than $\sim 80$ Myrs would not be expected for Type II SN~2017eaw-like progenitors (with final cores more massive than $4.4 ~\msun$), even taking into account binarity. Potentially a combination of other stellar assumptions (e.g., higher core mass and/or higher rotation) for intermediate mass stars that eventually experience a reverse merger to form the progenitor, could explain such a long age, but cannot be modelled with our setup up.

%------------------------
\subsubsection{Runaway Scenario} \label{sec:runaway}
%------------------------
One way to account for a discrepancy between the age of a SN and its  inferred final mass is with the hypothesis that the progenitor has moved away from its parent population over it's lifetime. 
Massive stellar progenitors may be kicked out of their parent population  through dynamical interactions in young stellar clusters \citep{Poveda1967} or due to a prior SN in a binary system \citep{Eldridge+2011,Renzo+2019}.

We investigate whether the progenitor of SN~2017eaw could have originated from the cluster seen in \autoref{fig:region_summary} (shown in red). 
Our fits without simultaneous fitting of a background CMD, considered out to 200 pc (the distance to the cluster), indicate the cluster age is $\sim$5 Myr.
This age implies the progenitor would had to have been moving $\geq$40 km s$^{-1}$ if it was expelled from the cluster, which is in the low-probability tail for a SN ejection \citep{Renzo+2019} but not unreasonable for a dynamical ejection \citep[e.g.,][]{Leonard1991}. % Not sure if it indeed so low 
However, the ejected star would have a maximum age of the cluster (5 Myr) which in all cases 
 would be too young to form a RSG and a Type II progenitor \citep[e.g.][]{Georgy+2012}.    Although we cannot exclude the possibility of a fine-tuned scenario where a binary system is dynamically ejected from the cluster \citep[e.g.][]{Perets+2012} and later merges, 
% consisting the dynamical ejection of a  binary system from the cluster \citep[e.g.][]{Perets+2012} and later merging and 
 producing a H-rich SN with a He core mass of of $4.4-6 ~\msun$ at ~5 Myrs, the scenario is considered extremely unlikely and as there is no further evidence for this, we do not focus on it. Thus, we conclude that the progenitor has probably not been ejected from the cluster.

There is still the possibility for an ejection of a companion from a prior SN in the field, and this was taken into account in the selection of 50, 100 and 150 pc radii of our analysis (\autoref{sec:reanalysis}), as they include the  typical distance of $\sim 100$ pc \citep{Renzo+2019} that runaway progenitors travel since the prior SN ejection.
%I need to go to ejecta masses, not envelope masses

%This is only for lower explosion energies, or nickel mass creation.

\subsection{ Ejecta masses}

\begin{figure}
\begin{center}
\includegraphics[width=3.6in]{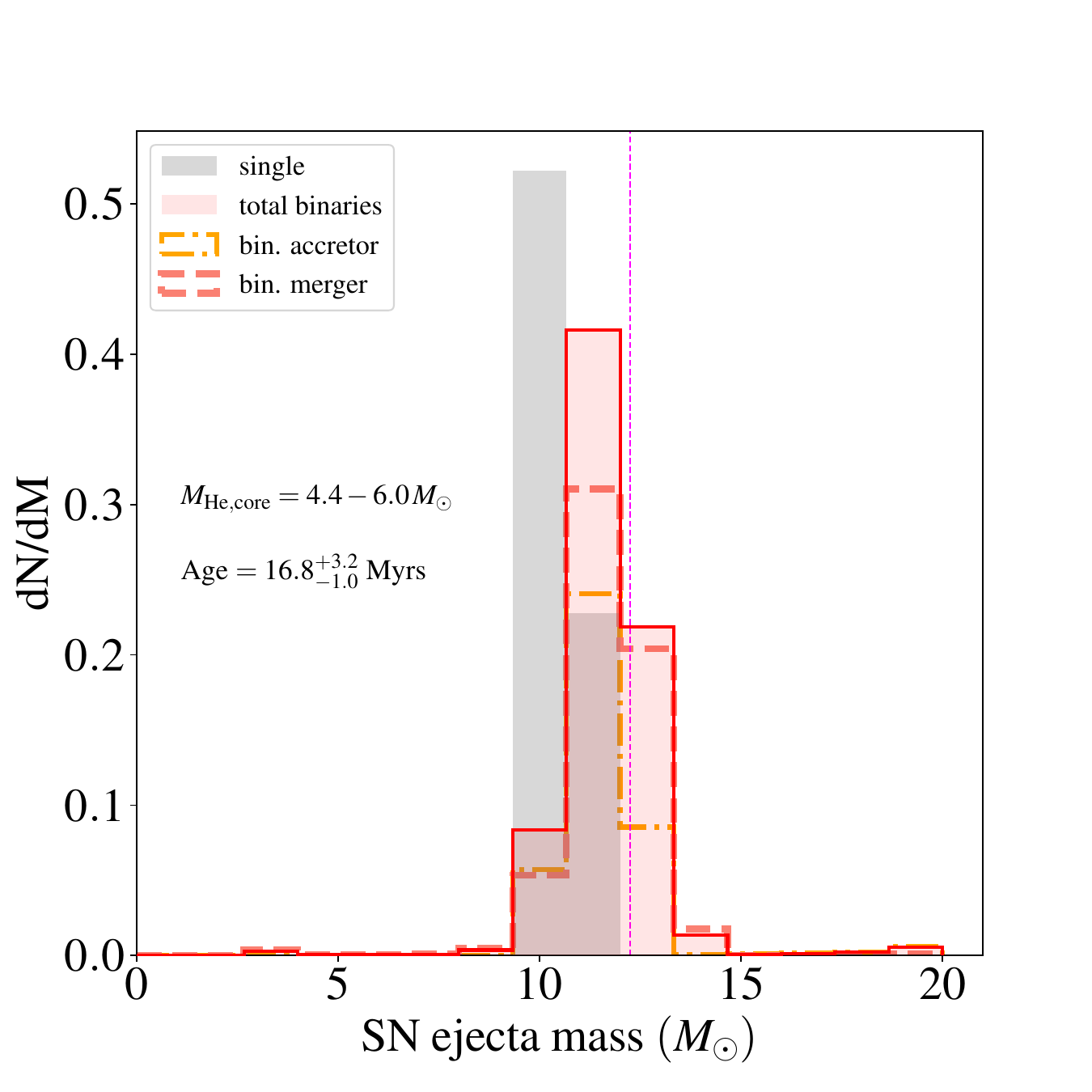}\\
 \caption{Normalized distribution of the estimated ejecta mass of Type II SN progenitors, with final helium core masses, $M_{\rm He,core} = 4.4 - 6.0 ~\msun$ 
 and an age estimate between $16.8^{+3.2}_{-1.0}$ Myrs, similar to SN~2017eaw-like events. We assume that a compact remnant mass of $1.4 ~\msun$ and the rest of the final, preSN mass is ejected. The normalized distribution of all binaries (red solid line) is subdivided into binary scenarios (mass gainers or MS mergers), and includes a broader spread of progenitors masses than the normalized distribution of single star progenitors. We also add the best fit inferred ejecta mass for SN2017eaw according to \citet{2020goldberg} (vertical dashed magenta line)
 \label{fig:ejecta_mass}}
  
\end{center}
\end{figure}

Our analysis is mostly focused on methods that probe final core mass and the derived age of the progenitor. In principle light curve method gives an extra constraint on the ejecta mass during the explosion, although this information is degenerate with the nickel mass produced, the explosion energy and the preSN progenitor's radius \citep{Goldberg+2019}. 

In Figure~\ref{fig:ejecta_mass} we show the estimated ejecta mass of the SN~2017eaw-like progenitors, where we define the ejecta mass as the final mass minus the compact remnant. We assumed a typical $1.4 ~\msun$ mass of a neutron star for the compact remnant, and that the rest of the final, preSN mass is ejected during the SN. We find that for single star progenitors, the ejecta mass is expected to be roughly $10 ~\msun$, originating from a single star of $13-15 ~\msun$ initial mass as shown in Figure~\ref{fig:initial_mass}. Binary progenitors show predominantly similar but slightly higher values of ejecta masses, due to gaining mass from their companion. For comparison we also we also show the $\sim 13.64 - 1.4 = 12.24 ~\msun$ ejecta mass inferred for 2017eaw from \citet{2020goldberg}, which at first glance is fairly consistent with both single and binary scenarios. 

Note that for stellar models, the uncertainty of the wind mass loss rate, especially during the RSG phase \citep[e.g.,][]{Smith2014,Beasor+2020,Yang+2023} limits the accuracy of the predicted final envelope (and thus ejecta) masses. In addition. for progenitors that experienced binary accretion, the poorly constrained mass transfer efficiency \citep[][]{de-Mink+2007} as well as the mass lost during the common-envelope phase \citep[e.g.][]{Ivanova+2013} for merger progenitors are extra sources of significant uncertainty. Given these caveats, we consider most of our single and binary models consistent with the inferred ejecta mass and refrain from opting in favour of a scenario. 
Nevertheless, we can still rule out some rare extreme binary channels with almost complete stripping during the merging phase or of very high accretion during the stable mass transfer of the progenitor. Still, ejecta masses can, in general, provide an important constraint for future studies, especially focusing on sub-classes of Type II SNe, such as short-plateau ones \citep{Eldridge+2019, 2021hiramatsu}.

%metallicity
\subsection{Metallicity}
 Our analysis was based on a stellar population of slightly sub-solar metallicity ($Z=0.008$), which is close but slightly lower than the inferred metallicity of the host region of SN~2017eaw, estimated around $Z=0.009-0.011$ \citep{2019vandyk}. Repeating our analysis for a simulation of solar-metallicity stellar population ($Z=0.0142$) resulted in similar qualitative conclusions. However, it is important to note that at solar metallicity, the age of single stars with our minimum assumed final core of 4.4 is lowered to $\sim 17$ Myrs, increasing the probability of SN~2017eaw being a binary progenitor even in the age window of $16.8^{+3.2}_{-1.0}$ Myrs above the 35\% limit of our default analysis. 
 As metallicity effects can be important for the application of this method to other SNe, especially in significantly sub-solar metallicity environments, 
 we show the correlation of age with final helium core mass, as in \autoref{fig:2D_age_Hecore}, for various metallicities in \autoref{sec:App_differentZ}.

%%%%%%%%%%%%%%%%%%%%%%%%%%%%%%%%%%%%%%%%%%
\section{Conclusion}\label{sec:conclusion}
%%%%%%%%%%%%%%%%%%%%%%%%%%%%%%%%%%%%%%%%%%
In this paper we have combined progenitor measurements from the stellar population at the SN site, pre-explosion imaging, and post-explosion imaging and spectroscopy. 
We place tighter constraints on the progenitor of SN~2017eaw and resolve a discrepancy in the literature between progenitor masses measured from observations near explosion and those that measure the initial progenitor mass.

From modeling the RSG progenitor prior to explosion, modeling the SN light curve, and modeling the late-time spectra of SN~2017eaw, a ZAMS progenitor mass between 12 and 15 $\msun{}$ was collected from the literature \citep{2019vandyk, 2018kilpatrick, 2019rui, 2020morozova, 2020goldberg, 2019szalai}. 
Additional progenitor mass constraints were provided from the literature analysis of SFH of the region around SN~2017eaw from \citet{2018williams,2021koplitz} who find  $M_{ZAMS}\leq 10.8$  $\msun$ ($M_{\rm He,core}\leq 3.4 ~\msun$)..
We then translated the ZAMS progenitor masses derived from observations near explosion to helium core masses resulting in a helium core mass range of $M_{\rm He,core} = 4.4 - 6.0 ~\msun$.

We revisit the analysis of the stellar population surrounding SN~2017eaw making a number of significant changes. 
We tune the quality cuts to this specific region, resulting in more stars in the CMD than previous analyses. 
We then examine the SFH with and without a background population fit simultaneously with the SN region.
While there continue to be no young populations when a separate background population is simultaneously fit with the SN population, when we use the full photometric catalog as the SN population, we find an age of $16.8^{+3.2}_{-1.0}$ Myr when a maximum age of 50 Myr is applied.  This age is confirmed using the recent UV SFH of this region by \cite{2023Tran}.

When an upper age limit of 150 Myrs, which is more  consistent with binary channels for SN progenitors, is applied, the age of the stellar population increases to $85.9^{+3.2}_{-6.5}$. 
This age is inconsistent with a single star  scenario for SN~2017eaw and is extreme even for the binary scenarios we consider for this event.

The previous results consider a region with a 50 pc radius centered on SN~2017eaw. We also explored the impact of expanding the radius of the region considered out to 200 pc, however, we find that beyond 150 pc, we include star clusters which drive the age of the population but are unlikely to be the birth population of SN~2017eaw, even if it had been kicked out of its natal cluster.

Finally, we combine the final helium core masses and the new population ages with single and binary population modeling to consider the single and binary progenitor channels allowed by our analysis.
We find the younger age of $16.8^{+3.2}_{-1.0}$ Myr has a 65\% chance of coming from a single star progenitor and a 35\% chance of having a prior binary mass accretion or merging phase, with the latter being  slightly more probable.  
For the most probable single star scenario, we find that the progenitor of SN~2017eaw originates from a $\sim 4.0-4.9~\msun$ helium core mass ($M_{ZAMS}\sim13-15~\msun$), which is consistent with those derived from other methods. 
When considering binary channels, a star that exploded with the same helium core mass could originate from a wider initial mass range of $\sim 11-16 ~\msun$.

While we consider it important to recognize the maximum age limit imposed by binary evolution (200 Myr) rather than single star evolution (50 Myr), in this case we find that the median stellar population when a maximum age of 150 Myr is used is inconsistent with any formation scenario. 
We encourage future works to consider both upper age limits.

Our analysis of the SN 2017eaw can be seen as a case study of identifying potential Type II SN progenitors with a history of binary mass accretion or merging and demonstrates the power of combining  different observational techniques that probe the initial and preSN properties of the progenitor.  
We encourage the community to use the theoretically expected correlation of age with helium core mass, shown in \autoref{fig:2D_age_Hecore}, as a framework applicable for identifications of potential binary progenitors of nearby Type II SN events for which a population age can be derived as well as a near explosion mass. As the correlation is slightly sensitive to the metallicity of the stellar population where the progenitor originated from, we provide the similar panels for different metallicities in the Appendix. 


\vspace{5mm}
\begin{acknowledgements}

We thank the referee for their careful reading of the paper and thoughtful comments.

Additional thanks to Jared Goldberg and Eric Bellm for fruitful discussions regarding the SN analysis in this paper and Leo Girardi for generating lookup tables between age and He core masses for stars $>$6$~\msun$.

E.Z. and K.A.B. acknowledge the organizers of IAUS 361 Massive Stars: Near and Far at which this paper was conceived of over a very productive breakfast.

E.Z. acknowledges funding support from the European Research Council (ERC) under the European Union's Horizon 2020 research and innovation programme (Grant agreement No. 772086),  as well as the Hellenic Foundation for Research and Innovation (H.F.R.I.) under the ``3rd Call for H.F.R.I. Research Projects to support Post-Doctoral Researchers'' (Project Number: 7933).

B.K. and B.F.W. acknowledge support for this work by NASA through grants GO-15216, GO-14786, and AR-15042 from the Space Telescope Science Institute, which is operated by AURA, Inc., under NASA contract NAS 5-26555. This work was supported by a grant from the Simons Foundation (CCA 928261, B.F.W.).

K.A.B. is supported by an LSSTC Catalyst Fellowship; this publication was thus made possible through the support of Grant 62192 from the John Templeton Foundation to LSSTC. The opinions expressed in this publication are those of the authors and do not necessarily reflect the views of LSSTC or the John Templeton Foundation.
\end{acknowledgements}
\facilities{}


\software{astropy \citep{2018AJ....156..123A},  
          binary population synthesis code {\tt binary\_c} \citep{Izzard+2004,Izzard+2006,Izzard+2009}, KEPLER \citep{1978weaver}, MATCH \citep{2002dolphin,2012dolphin,2013dolphin}, MESA \citep{2011paxton, 2013paxton, 2015paxton, paxton_modules_2018, paxton_modules_2019}, SNEC \citep{2015morozova}, STELLA \citep{1998blinnikov, 2000blinnikov, 2006blinnikov} }



\appendix

%\section{Appendix information}

%\begin{figure*}
%    \centering
%    \includegraphics[width=0.85\textwidth]{SFHs_50Myr_log_0107.png}
%    \caption{Same as Figure 2 but in log age rather than Myr.}
%\end{figure*}

\section {Age vs core mass for different metallicities} \label{sec:App_differentZ}
To facilitate the application of this method to other SNe, in potentially different metallicity environments, we provide \autoref{fig:2D_age_Hecore} for a range of metallicities ($Z=0.0002-0.02$).  For lower metallicities the lifetime of the Type II progenitors, both singles and binaries, is longer. This is predominantly due to the fact that at lower-Z, star of fixed final core mass originate from lower initial masses, with longer evolutionary timescale \citep{Pols+1998}. Also for $Z=0.02$ we do not find Type II SN progenitors with final helium-core masses higher than $\sim 8.5~\msun$, as the enhanced wind mass loss prevents progenitors to form more massive cores.
\begin{figure*}
    \centering
    \includegraphics[width=0.45\textwidth]{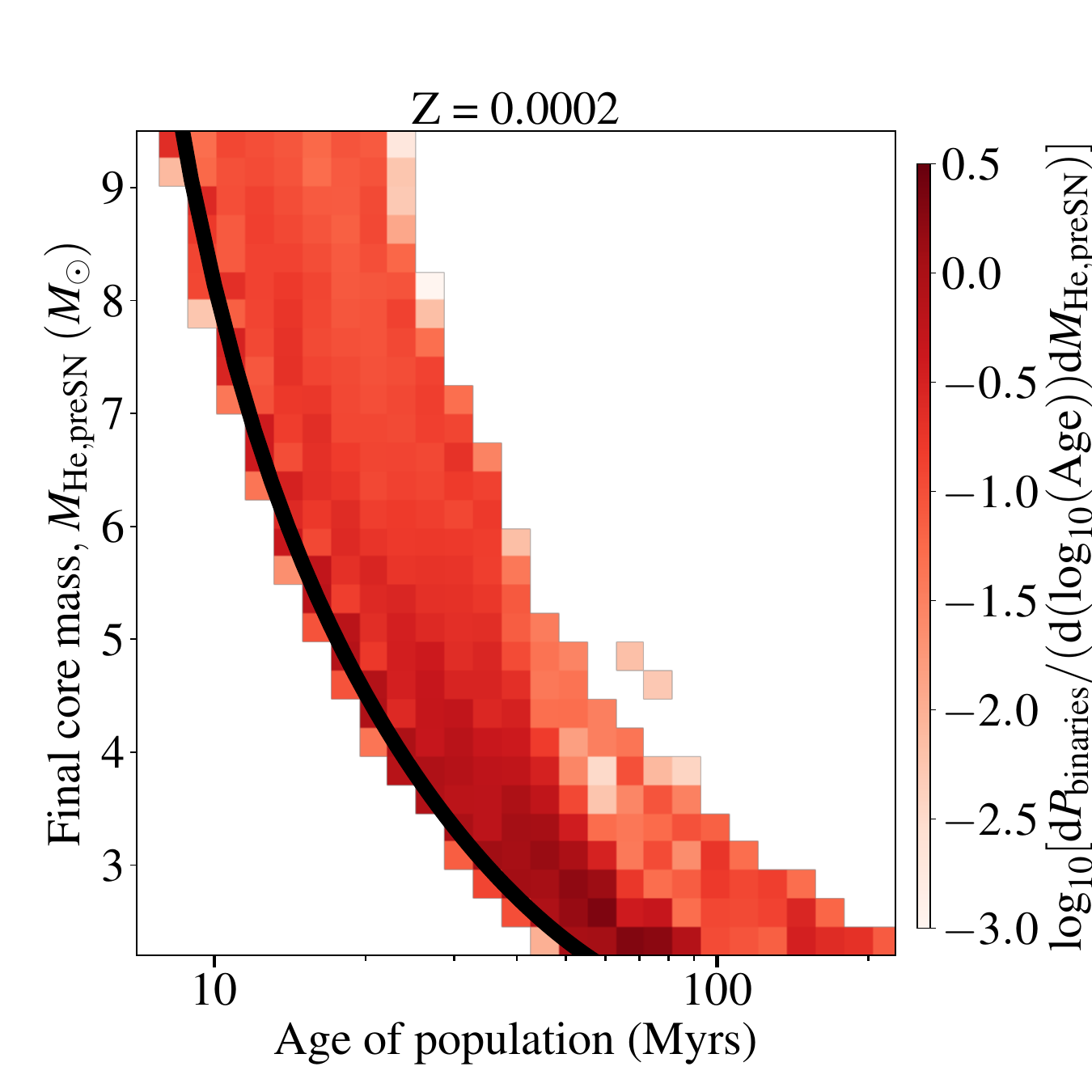}\includegraphics[width=0.45\textwidth]{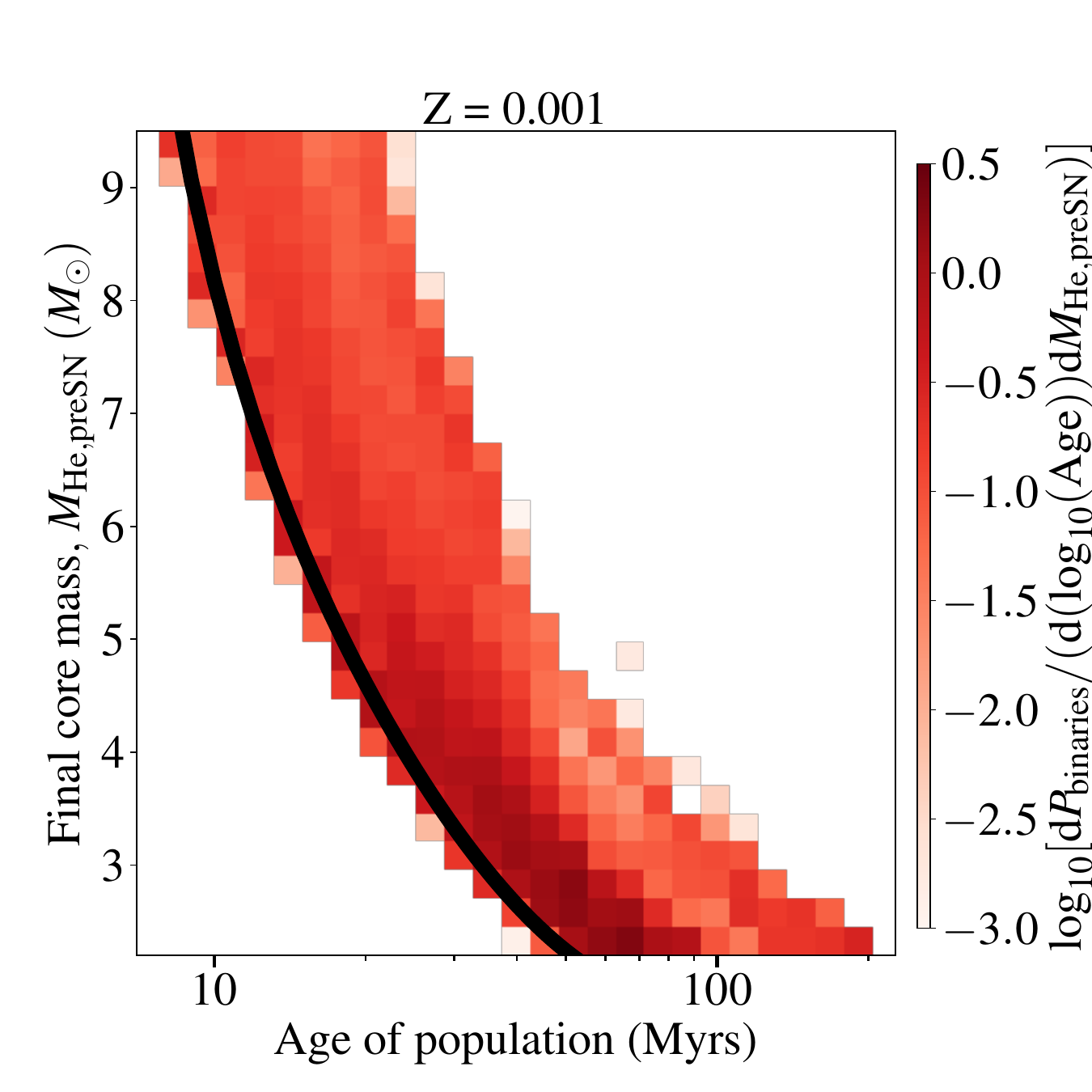}\\
    \includegraphics[width=0.45\textwidth]{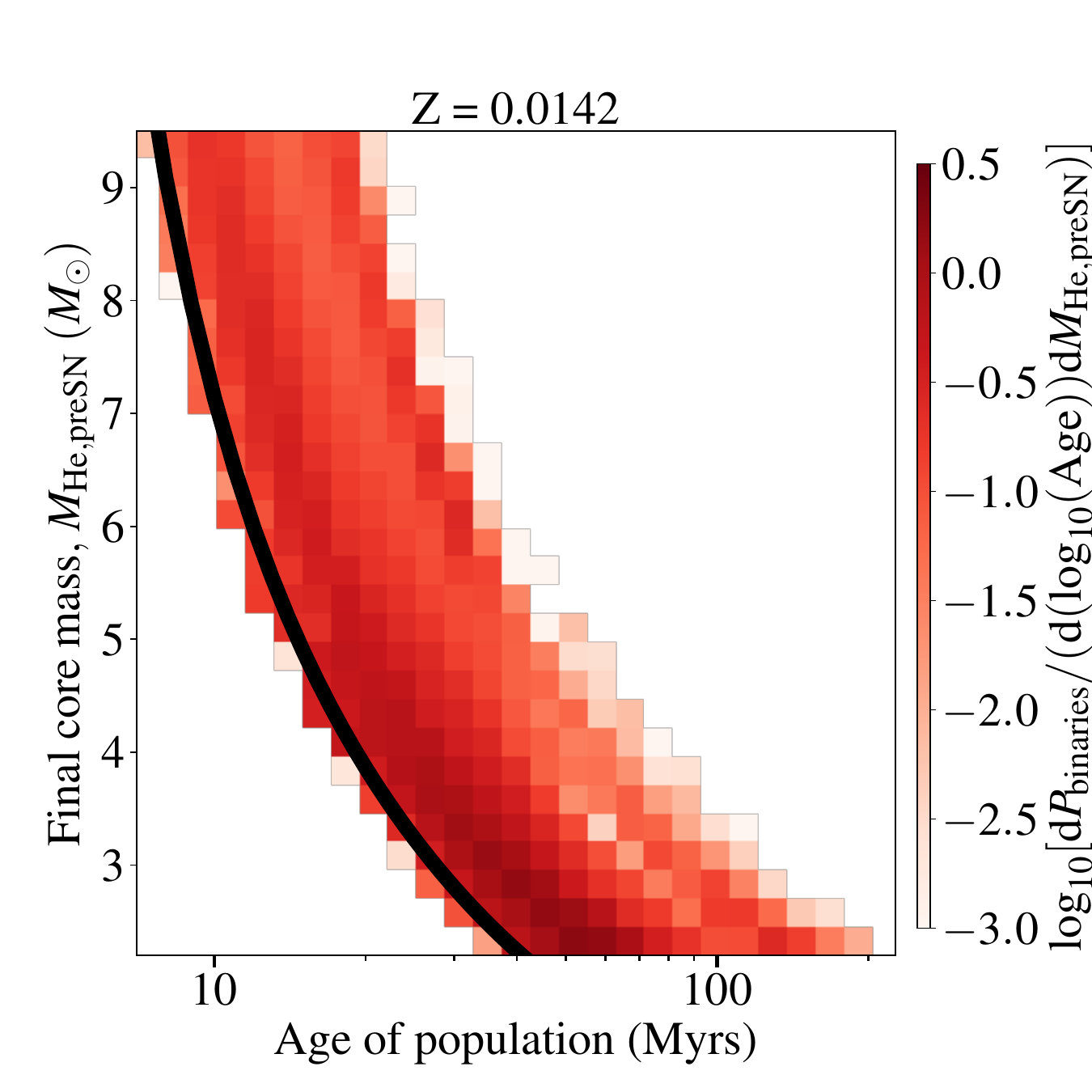}
    \includegraphics[width=0.45\textwidth]{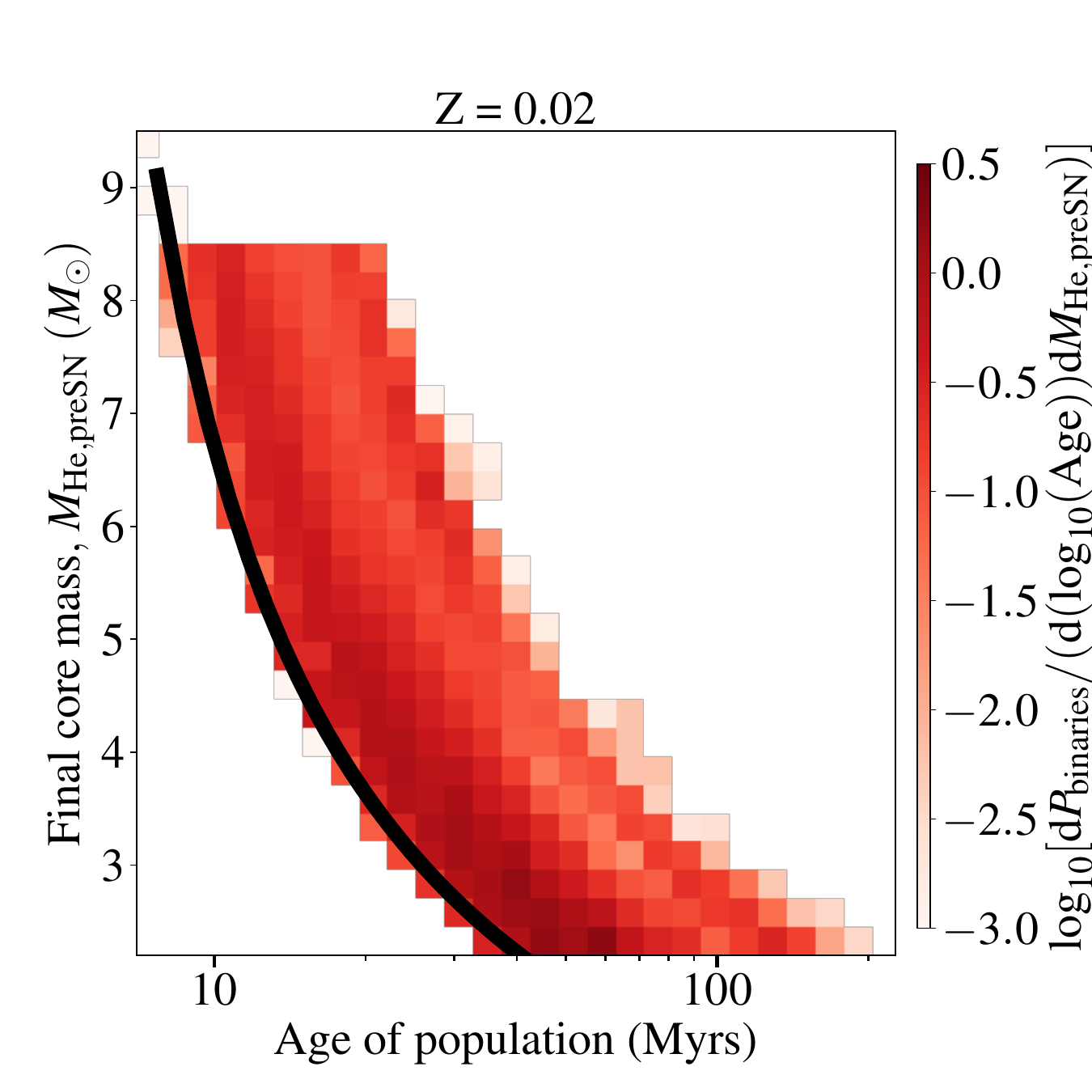}
    \caption{Same as \autoref{fig:2D_age_Hecore}, for different metallicities. Lower metallicity shifts the distribution towards longer ages. For Z=0.02, due to stronger stellar winds, there are no giant Type II progenitors with final core mass above $\sim 8.5  ~\msun$.}
\end{figure*}

%\begin{figure}
%    \centering
%    \includegraphics[width=0.90\textwidth]{prob_typeII_constrainedmass4_Z0.008_withBrads_Ageprobability.pdf}
%    \caption{Alternative to Figure 5}
%\end{figure}

\bibliography{bibliography}{}
\bibliographystyle{aasjournal}



\end{document}